\documentclass[rmp, aps, nofootinbib,11pt, letterpaper]{revtex4}
\usepackage{epsfig}
\usepackage{amsmath,amssymb}
\usepackage{fancyhdr}
\usepackage[normalem]{ulem}
\usepackage{color}
\usepackage{mathrsfs}  
\usepackage{hanging}
\usepackage[breaklinks=true,colorlinks=true, linkcolor=blue, citecolor=magenta]{hyperref}
\usepackage{url}

\usepackage{pifont}

\usepackage{float, placeins}
\usepackage{mwe}
\usepackage[caption=false]{subfig}
\usepackage{morefloats}
\usepackage{empheq}
\usepackage{bm}
\usepackage{booktabs} 
\usepackage{multirow}
\usepackage{makecell}
\usepackage{cases}
\usepackage[normalem]{ulem}
\usepackage[margin= 1in]{geometry}
\usepackage{setspace}

\makeatletter
\gdef\@ptsize{1}
\makeatother

\allowdisplaybreaks

\newcommand{\nc}{\newcommand}
\nc{\ba}{\begin{eqnarray}}
\nc{\ea}{\end{eqnarray}}
\newcommand\be{\begin{equation}}
\newcommand\ee{\end{equation}}

\def\Pinf{P_{\mathrm{inferred}}}
\def\PrA{Pr_A}
\def\PrD{Pr_D}

\begin{document}

\title{{\large A Bayesian View on the Dr.\ Evil Scenario}}

\author{Feraz Azhar}
\email[Email address: {fazhar@nd.edu}]{}
\affiliation{Department of Philosophy, University of Notre Dame, Notre Dame, IN 46556, USA\\\vspace{-0.4cm}%
and Black Hole Initiative, Harvard University, Cambridge, Massachusetts 02138, USA}

\author{Alan H. Guth}
\email[Email address: {guth@ctp.mit.edu}]{}
\affiliation{Department of Physics, Laboratory for Nuclear
Science, and Center for Theoretical Physics, Massachusetts Institute of Technology, Cambridge, MA 02139}

\author{\vskip -10pt Mohammad Hossein Namjoo}
\email[Email address: {mh.namjoo@ipm.ir}]{}
\affiliation{School of Astronomy,
	Institute for Research in Fundamental Sciences (IPM), Tehran, Iran}

\date{{December 19, 2022}}

\begin{abstract}
\vspace{-1cm}
\singlespacing
In {\it Defeating Dr.~Evil with Self-Locating Belief}, Adam Elga proposes and defends a principle of indifference for self-locating beliefs: if an individual is confident that his world contains more than one individual who is in a state subjectively indistinguishable from his own, then he should assign equal credences to the hypotheses that he is any one of these individuals.  Through a sequence of thought experiments, Elga in effect claims that he can derive the credence function that should apply in such situations, thus justifying his principle of indifference.  Here we argue, using a Bayesian approach, that Elga's reasoning is circular: in analyzing the third of his thought experiments, he uses an assertion that is justifiable only if one assumes, from the start, the principle of indifference that he is attempting to justify.  We agree with Elga that the assumption of equal credences is a very reasonable principle, in the absence of any reason to assign unequal credences, but we do not agree that the equality of credences can be so derived.

\end{abstract}

\maketitle
\vspace{-1cm}
\singlespacing
\tableofcontents
\singlespacing

\section{Introduction}\label{SEC:Intro}

Self-locating beliefs---namely, those beliefs that situate an agent at a location or a time---are commonplace. You acquire a self-locating belief whenever you come to learn of the time by glancing at your watch or if you learn of your location by observing a street sign. The converse situation can also arise, in which you may be initially certain about some self-locating belief (such as the time) but then you gain a new piece of evidence (that your watch is broken), rendering you uncertain about the self-locating belief~\cite{bradley_07}. More generally, the question arises as to how one should distribute one's credence over self-locating beliefs about which one is uncertain.

In {\it Defeating Dr.~Evil with Self-Locating Belief}, \citet{elga_04} defends a response to this question. The response amounts to a version of the principle of indifference---where, under specified circumstances, one should spread one's credence equally over the various hypotheses. More specifically, the version of the principle of indifference that Elga defends is as follows:
\begin{quote}
\textsc{Indifference}: Similar centered worlds deserve equal credence~\cite[p.~387]{elga_04}.
\end{quote}
A centered world is a possible world with a designated individual and a designated time. It can be specified as a triple $(w; i; t)$, where $w$ is a possible world, $i$ is some individual in $w$, and $t$ is a time. Similar centered worlds satisfy two conditions: (i) they agree on the first argument of the triple (but not necessarily on the latter two); (ii) the individuals that exist in each centered world are in subjectively indistinguishable states.\footnote{\textsc{Indifference} can be contrasted with the claim that two triples in which the possible worlds themselves are different (but where the individuals are in subjectively indistinguishable states) should receive equal credence---a claim that Elga holds to be absurd, since two possible worlds can have very different levels of plausibility.} Note that while principles of indifference have been extensively discussed, the formulation in~\citet{elga_04} has been particularly influential.\footnote{For example, Bradley devotes an entire chapter of his Ph.D. thesis [\citet{bradley_07}] and an entire section of a paper [\citet{bradley_11}] to defending Elga's formulation of indifference against criticisms by~\citet{weatherson_05}. \citet{birch_13}---in his refutation of Bostrom's (\citeyear{bostrom_03}) argument that we may very well be living in a computer simulation---explores Elga's formulation of indifference as a possible justification of Bostrom's assumptions. \citet{wilson_17} describes the applicability of the principle of indifference in Everettian quantum mechanics, employing Elga's formulation throughout.  \citet{sebens+carroll_18} view their work as applying Elga's principle of indifference to Everettian quantum mechanics to show that it leads to the Born rule for probabilities of the outcomes of measurements. And the {\it proof} that Elga provides for his formulation of the principle of indifference has also been persuasive.~\citet[p.~4]{carroll+sebens_15} declare that %
\begin{quote}
Elga \dots has given convincing arguments in favor of indifference in the case of identical classical observers. Crucially, this result is not simply postulated as the simplest approach to the problem, but rather derived from seemingly innocuous principles of rational reasoning.
\end{quote}
}

In justifying \textsc{Indifference}, Elga considers a sequence of three (related) thought experiments: \textsc{Duplication, Toss\&Duplication}, and \textsc{Coma}.  We claim that Elga's analysis of the \textsc{Coma} scenario is flawed, relying on an assertion which can be justified only by assuming the truth of the proposition that he is trying to demonstrate.  We are not disputing the claim that \textsc{Indifference} is reasonable---indeed, in an upcoming paper we will make use of such a principle in assessing certain cosmological theories.  Our remarks here target only the justification that Elga provides for \textsc{Indifference}.  For us, \textsc{Indifference} is a principle that one can reasonably adopt, based on some sort of appeal to the absence of any reason to assign unequal credences, but not based on any determinative calculation of the sort presented by Elga.\footnote{Since we agree that \textsc{Indifference} is a reasonable principle, we also agree with Elga's conclusion that Dr. Evil ought to surrender, assuming that the fear of torture outweighs the thrill of his evil plans!}

Our plan for this paper is as follows. In Sec.~\ref{SEC:Evidence} we summarize Elga's thought experiments, and describe our claim that the analysis of \textsc{Coma} is flawed. Sec.~\ref{SEC:Bayes} describes a Bayesian calculation that shows in detail how we believe the \textsc{Coma} scenario should be analyzed. We summarize our argument in Sec.~\ref{SEC:Conclusion}. In an appendix, we reexamine step-by-step a crucial footnote from Elga's paper, showing how the conclusions found there are modified by our analysis of \textsc{Coma}.

\section{Reanalysis of Elga's thought experiments}\label{SEC:Evidence}

Elga's first thought experiment describes two similar centered worlds in which a person named ``Al'' is duplicated. The experiment is described as follows.
\begin{quote}
\textsc{Duplication}: After Al goes to sleep researchers create a duplicate of him in a duplicate environment. The next morning, Al and the duplicate awaken in subjectively indistinguishable states~\cite[p.~388]{elga_04}.
\end{quote}
The issue at stake is how Al should distribute his credence between the hypothesis that he is Al and the hypothesis that he is the duplicate. In order to justify why Al should distribute his credence {\it evenly} between the two hypotheses (which is indeed what would be implied by \textsc{Indifference}), Elga introduces two further experiments: \textsc{Toss\&Duplication} and \textsc{Coma} (the latter will be described shortly).
\begin{quote}
\textsc{Toss\&Duplication}: After Al goes to sleep, researchers toss a coin that has a 10\% chance of landing heads. Then (regardless of the toss outcome) they duplicate Al. The next morning, Al and the duplicate awaken in subjectively indistinguishable states~\cite[p.~388]{elga_04}.
\end{quote}
In our discussion we will generalize the chance of the coin landing heads from the specified value of 10\% to an arbitrary $P_0(H)$, assuming only that $P_0(H)$ is not equal to 0 or 1. Elga claims (and we agree) that
\begin{quote}\dots Al's state of opinion (when he awakens) as to whether he is Al or the duplicate ought to be the same in \textsc{Toss\&Duplication} as it is in \textsc{Duplication}. So in order to show that in \textsc{Duplication}, Al ought to divide his credence evenly between the hypothesis that he is Al and the hypothesis that he is the duplicate, it is enough to show that he ought to do so in \textsc{Toss\&Duplication}~\cite[pp.~388--389]{elga_04}.
\end{quote}

To show that Al ought to divide his credence evenly between the hypothesis that he is Al and that he is the duplicate (in \textsc{Toss\&Duplication}), Elga states three claims, which we (re)describe here. Following \citet{elga_04} and \citet{weatherson_05}, we will use the following abbreviations:
\begin{align}
H:&\textrm{ the coin lands `heads'};\nonumber\\ T:&\textrm{ the coin lands `tails'};\nonumber\\ A:&\textrm{ I am Al};\nonumber\\ D:&\textrm{ I am Dup (Al's duplicate)}.\nonumber
\end{align}
The credence function that Al ought to have immediately upon awakening will be denoted by $P(\cdot)$.  Elga's three claims can be stated as follows:
\begin{quote}
\begin{itemize}
\item[(C1)] Al's credence in $H$ ought to be equal to the chance of the coin landing heads:
\ba
P(H)=P_0(H). \label{EQN:C1}
\ea
\item[(C2)] Al's credence in $H$, given (($H$ and $A$) or ($T$ and $A$)), ought to be equal to his credence in $H$:
\ba
P(H | HA \hbox{ or } TA) = P_0 (H). \label{EQN:C2}
\ea
\item[(C3)] Al's credence in $H$, given (($H$ and $A$) or ($T$ and $D$)), ought to be equal to his credence in $H$:
\ba
P(H | HA \hbox{ or } TD) = P_0(H). \label{EQN:C3}
\ea
\end{itemize}
\end{quote}

Claim (C3) (which Elga indeed deems to be controversial) is established by considering another thought experiment, viz.~\textsc{Coma}.
\begin{quote}
\textsc{Coma}: As in \textsc{Toss\&Duplication}, the experimenters toss a coin and duplicate Al. But the following morning, the experimenters ensure that {\it only one person wakes up}: If the coin lands heads, they allow Al to wake up (and put the duplicate into a coma); if the coin lands tails, they allow the duplicate to wake up (and put Al into a coma)~\cite[p.~390--391]{elga_04}.
\end{quote}

Elga then claims (a claim with which we agree) that one can determine the value of the left-hand side of Eq.~\eqref{EQN:C3} by considering what Al's credence in $H$ should be in \textsc{Coma}.  That is, when Al awakens in \textsc{Coma}, were he to indeed awaken, his situation would be exactly as it would have been in \textsc{Toss\&Duplication}, but then updated by the new information $(HA \hbox{ or } TD)$. Thus, Al's credence function in \textsc{Coma}, if he awakens, ought to be given by
\begin{equation}\label{EQN:Pcomagen}
P_{\textsc{Coma}}(\cdot) = P(\cdot|HA \hbox{ or } TD).
\end{equation}
In particular,
\begin{equation}\label{EQN:Pcoma}
P_{\textsc{Coma}}(H) = P(H|HA \hbox{ or } TD).
\end{equation}%

Al's credence in $H$ in \textsc{Coma} is ascertained, by Elga, using the following argument (on p.~392)\footnote{\label{fn:ExactQuote} Here, we have modified the exact quote to accord with our notation, changing ``HEADS'' to $H$ and ``10\%'' to $P_0(H)$.}:
\begin{quote}
Before Al was put to sleep, he was sure that the {\it chance} of the coin landing heads was $P_0(H)$, and his credence in $H$ should have accorded with this chance: it too should have been $P_0(H)$.  When he wakes up, his epistemic situation with respect to the coin is just the same as it was before he went to sleep. He has neither gained nor lost information relevant to the toss outcome. So his degree of belief in $H$ should continue to accord with the chance of $H$ at the time of the toss. In other words, his degree of belief in $H$ should continue to be $P_0(H)$.
\end{quote}
In short, Elga's claim is that in \textsc{Coma}, if Al awakens, he ought to continue to set his credence in the coin landing heads to the known chance of heads, $P_0(H)$. 

Given claims (C1), (C2), and (C3), Elga (correctly) concludes that $P(A)=P(D)=1/2$ in \textsc{Toss\&Duplication} and thus (by the reasoning described above) in \textsc{Duplication}. We, however, take issue with the justification provided for claim (C3).

From our point of view, Elga appears to be using circular reasoning in (C3), where he claims that Al's credence in $H$ should remain equal to $P_0(H)$ when he awakens in \textsc{Coma}. In particular, it is not the case that when Al wakes up ``his epistemic situation with respect to the coin toss is just the same as it was before he went to sleep. He has neither gained nor lost information relevant to the toss outcome.'' We contend that Al has both gained and lost information relevant to the toss outcome:
\begin{itemize}
\item[(i)] He has lost information as regards his identity:
before Al went to sleep he was sure that he was Al; after he awakens, the possibility arises that he is Dup. No amount of introspection (or an inspection of his external environment) can reveal to him whether he is Al or Dup. (This is what we take to be the meaning of the assumption that Al and Dup would be in ``subjectively indistinguishable'' states.)
\item[(ii)] He has gained the information that he is now in a predicament in which his identity (again, about which he is now unsure) is perfectly correlated with the toss outcome. 
\end{itemize}
Al's uncertainty about his identity [as described in (i)] is directly relevant to his belief about the toss outcome [as described in (ii)]. 

In particular, when Al awakens in \textsc{Toss\&Duplication}, he has no way of knowing if he is Al or Dup.  If he adopts \textsc{Indifference}, he will conclude that he should give equal credence to each possibility.  However, since Elga is trying to demonstrate \textsc{Indifference}, circularity can be avoided only if we allow Al to adopt an initial credence that is not equal to 1/2.  Hence we set $P(A)$, Al's credence that he is Al upon awakening in \textsc{Toss\&Duplication}, equal to some arbitrary `prior' credence $\PrA$, which Al is free to choose.%
\footnote{For some background on `prior' credence functions, see for example~\citet{meacham_16},~\citet{dorr+arntzenius_17}, and~\citet{isaacs+al_21}.  In contrast to some ways of understanding such a prior
credence function, in the present setting $P(\cdot)$ is not to be understood as the credence function of an agent who has no evidence whatsoever. Al, for example, does know (among other things) that he is in a scenario in which he has been duplicated.}

In \textsc{Coma}, as discussed just above Eq.~(\ref{EQN:Pcomagen}), Al (if and when he awakens) begins with the same credences as in \textsc{Toss\&Duplication}, which are then updated by the new information ($HA \hbox{ or } TD$). The role of $\PrA$ is important, because if $\PrA$ is not equal to 1/2, then clearly Al's credence in $H$ {\it is} affected by the new information. For example, if $\PrA$ is nearly one, then his credence in $H$ should obviously increase on awakening in \textsc{Coma}, since if he is Al, then the coin must have landed heads. This issue is well-described by a standard Bayesian analysis, which we give in the next section.  We will see that Al's credence in $H$ upon awakening in \textsc{Coma} should remain $P_0(H)$ if and only if Al assumes that $\PrA = 1/2$. This means that Elga's conclusion is valid if and only if one assumes \textsc{Indifference} from the start.%
\footnote{Note that \citet{weatherson_05} has also objected to Elga's claim (C3), but makes no mention of circularity or anything similar.  His primary objection (although he raises others as well) relies on the view that one must distinguish between {\it risky} propositions, for which ``we have good reason to assign a particular probability,'' and propositions which are {\it uncertain}, for which we ``aren't really in a position to assign anything like a precise numerical probability.'' Weatherson argues that Al's question about his identity falls in the category of uncertainty.  The result of the coin toss was risky before Al went to sleep, but when he awakens in the \textsc{Coma} scenario, the result of the coin toss becomes correlated with Al's identity, which changes it from risky to uncertain.  Thus Weatherson questions whether Al can assign any credence, when he awakens, to the coin having landed heads, and suggests that maybe assigning a range of credences would be more appropriate.}

\section{A Bayesian approach}\label{SEC:Bayes}

In this section we apply Bayes' theorem to determine how Al should update his credence in $H$, using the new information, ($HA \hbox{ or } TD$), that he acquires on awakening in the \textsc{Coma} scenario.  To determine how Al should update his credence in $H$, we will compute the right-hand side of Eq.~(\ref{EQN:Pcoma}), recalling that $P(\cdot)$ is the credence function in \textsc{Toss\&Duplication}.

Our derivation will make use of the fact that in \textsc{Toss\&Duplication} the $H/T$ choice is independent of the $A/D$ choice, which follows as a consequence of claim (C2) (with which we do not take issue). This independence follows from the same words that Elga uses to justify claim (C2): ``So Al should count the toss outcome as irrelevant to who he is'' \cite[p.~389]{elga_04}. Formally, the independence of $H$ and $A$ can be derived from claim (C2) by noting that $(H \textrm{ or } T)$ is true, so $P(H | HA \hbox{ or } TA) = P(H | A)$, and therefore (C2) implies that $P(H | A) = P_0(H)$. This is of course a way of stating that $A$ and $H$ are independent.  Using Bayes' theorem, this statement is equivalent to $P(A|H) = \PrA$.

Recalling that $P(H) = P_0(H)$ is the initially specified chance that the coin landed heads, Bayes' theorem can be used to rewrite the right-hand side of Eq.~(\ref{EQN:Pcoma}) as follows:
\begin{subequations}
\label{EQ:Indep0}
\begin{align}
P(H | HA \hbox{ or } TD) &=\frac{P(HA \hbox{ or } TD|H)}{P(HA \hbox{ or } TD)}P_0(H)\\
     &=\frac{P(A|H)}{P(HA)+P(TD)}P_0(H)\\
     &=\frac{\PrA}{P_0(H){\PrA}+P_0(T)\PrD}P_0(H)
				   \label{EQ:Indep1}\\
     &\equiv F \, P_0(H),
\end{align}
\end{subequations}
where $\PrD \equiv 1-\PrA$ is Al's prior credence in \textsc{Toss\&Duplication} that he is Dup, and
\ba\label{EQN:F}
F \equiv \frac{\PrA}{P_0(H) {\PrA} + P_0(T) \PrD}.
\ea
In Eq.~\eqref{EQ:Indep1} we have used the independence of $H$ and $A$ (and that of $T$ and $D$).  As we claimed at the end of the previous section, Al's credence in $H$ after awakening in \textsc{Coma}, which is equal to $P(H | HA \hbox{ or } TD)$, remains equal to $P_0(H)$ only if $F=1$.  By rewriting $F$ as
\ba
F = \dfrac{1}{1 + \dfrac{P_0(T)}{\PrA} \bigl[{\PrD} - {\PrA}\bigr]},
\ea
one can easily see that $F=1$ only if the prior credences ${\PrA}$ and ${\PrD}$ (which must sum to 1) are each taken to be 1/2.\footnote{Recall that we have assumed that $P_0(H)$ is not equal to zero or one.  If we had allowed $P_0(H)=0$, then $F P_0(H)$ would equal $P_0(H)$ for any finite value of $F$.  If we had allowed $P_0(H)=1$ (and hence $P_0(T)=0$), then Eq.~(\ref{EQN:F}) shows that $F$ would equal 1 for any nonzero value of $\PrA$.}

Thus, taking into account the implications of Bayes' theorem, claim (C3) should be modified to a new claim, which we will denote by (C$3^{\prime}$):
\begin{quote}
\begin{itemize}
\item[(C$3^{\prime}$)] Al's credence in $H$ 
if he awakens in \textsc{Coma}, $P_{\textsc{Coma}}(H)$, which is equal to $P(H | HA \hbox{ or } TD)$, ought to be equal to the product of $F$ [given in Eq.~\eqref{EQN:F}] and his credence in heads $P(H)=P_0(H)$ as he awakens, before he updates his credences with the new information ($HA \hbox{ or } TD$):
\ba
P_{\textsc{Coma}}(H) =
P(H | HA \hbox{ or } TD) = F \, P_0(H). \label{EQN:C3prime}
\ea
\end{itemize}
\end{quote}
Note that (C$3^{\prime}$) agrees with (C3) only if ${\PrA}$ is assumed to be 1/2.

In his footnote 8, Elga uses claims (C1), (C2), and (C3) to show that $P(A)=P(D)$, which completes his demonstration that \textsc{Indifference} is true for this situation.  (In an appendix, he generalizes the argument to defend arbitrary instances of \textsc{Indifference}.) Since we have argued that (C3) should be replaced by (C$3^{\prime}$), the conclusion will of course be altered.  In deriving Eq.~\eqref{EQN:C3prime}, we calculated $P(H | HA \hbox{ or } TD)$, and hence $F$ [see Eqs.~\eqref{EQ:Indep0} and \eqref{EQN:F}], assuming that ${\PrA}$ and $P_0(H)$ were given. We could imagine, however, that we were given the values of $F$ and $P_0(H)$, and were asked to infer the value of $P(A)$, Al's credence in being Al when he awakens in \textsc{Toss\&Duplication}.  Equation~\eqref{EQN:F} would still hold [replacing $\PrA$ and $\PrD$ by generic quantities $P(A)$ and $P(D)$], so we could then find $P(A)$ directly from Eq.~\eqref{EQN:F}, recalling that $P(D) = 1 - P(A)$ and $P_0(T) = 1 - P_0(H)$. The result would be
\ba \label{Eq:PAd}
\Pinf(A)
=\dfrac{F P_0(T)}{1 - F + 2 F P_0(T)},
\ea
where we use the special notation $\Pinf(A)$ to denote the value of $P(A)$ that is inferred from assumptions (C1), (C2), and (C$3^{\prime}$).  This calculation is a version of Elga's derivation in footnote 8, which is shorter than Elga's argument and allows an arbitrary value for $F$.  We can see immediately that if we assume that $F=1$, as in claim (C3), we recover Elga's result that $\Pinf(A) = 1/2$.  However, if we use the version of the claim based on Bayesian reasoning, namely (C$3^{\prime}$), we find instead [by substituting Eq.~\eqref{EQN:F} into Eq.~\eqref{Eq:PAd}] the trivial conclusion that
\ba \label{Eq:Circ}
\Pinf(A)={\PrA}.
\ea 
That is, Al's credence in being Al upon awakening in \textsc{Toss\&Duplication}, as inferred by considering the experiment in \textsc{Coma}, is exactly equal to whatever prior credence ${\PrA}$ that he assumed. Al will conclude that $\Pinf(A) = 1/2$ only if he assumed a prior credence ${\PrA}$ equal to 1/2.

The argument above is an abbreviated version of Elga's footnote 8, modified to use (C$3^{\prime}$) instead of (C3).  In an appendix we show in detail that if the steps of Elga's footnote 8 are followed exactly, but where $F$ is introduced as in Eq.~\eqref{EQN:C3prime}, we retrieve exactly the result of Eq.~\eqref{Eq:PAd}.  Thus we see that once the circularity in Elga's formulation is removed, the demonstration of the truth of \textsc{Indifference} disappears. 
\textsc{Indifference}
can still be adopted as a reasonable principle, but Elga's derivation of it is flawed.

\section{Conclusion}\label{SEC:Conclusion}

\citet{elga_04} discusses a specific version of the principle of indifference, which applies to a situation where a possible world includes two or more individuals who, at some specified time for each of them, are in subjectively indistinguishable states.  He illustrates this situation with an example called \textsc{Duplication}, in which someone named Al is duplicated while he sleeps, along with his environment, so that Al and his duplicate awaken in subjectively indistinguishable states.  Through a sequence of three thought experiments, Elga argues that when Al awakens, he ought to assign equal credence to being the duplicate or to being Al.

In this paper we have argued that, while it is perfectly reasonable for Al to {\it assume} that he is equally likely to be the duplicate, he is not compelled to make this assumption.  The reasoning that Elga used, we believe, is circular.  Specifically, we differ in the analysis of the third thought experiment, called \textsc{Coma}, in which a coin with a 10\% chance of landing heads is tossed while Al sleeps, and then the duplication takes place as before.  If the coin lands heads, only Al is allowed to wake up, with the duplicate remaining in a coma; but if the coin lands tails, only the duplicate is allowed to wake up.  Our disagreement centers on the credence that Al should have, when (and if) he awakens, in the coin having landed heads.  Elga's conclusions are based on the claim that Al's credence in heads when he awakens should remain 10\%.  We argue, however, that this claim is true if and only if Al {\it assumes} that he is equally likely to be Al or the duplicate, which is exactly the conclusion that Elga is trying to demonstrate.  If Al does not make this assumption, then he might, for example, assume that he is much more likely to be Al than the duplicate.  In that case, his credence in heads should be increased upon learning that he has woken up.  We carried out a Bayesian analysis of this thought experiment, and showed that Al may assume any prior credence in his being Al, and no inconsistencies arise.

As long as Al has no reason to believe that he is more likely to be either Al or the duplicate, then we agree that it is reasonable to quantify this absence of evidence by adopting the default proposition that he is equally likely to be either.  Elga's argument, however, did not rely on adopting a default proposition.  Instead, Elga claimed to show directly from the descriptions of the thought experiments that Al could {\it deduce} that he should have equal credence in being Al or Dup. At the end of his argument, Elga proclaimed ``So, \textsc{Indifference} is true.''  Thus, Elga was arguing that \textsc{Indifference} is more than a reasonable proposition, but is instead a logically compelling conclusion.  We maintain, however, that this argument is flawed.

In an email exchange with Adam Elga, he pointed out that circularity could be avoided by accepting (C3) as an ``undefended premise in the argument.''  To ensure the absence of circularity, it is important that (C3) has ``independent appeal---credibility that does not derive from an antecedent commitment to \textsc{Indifference}.''  We completely agree that if one does not provide support for (C3), but instead accepts it as a premise, then there is no circularity.  For Elga, (C3) has appeal that is independent of \textsc{Indifference}.  For us, however, (C3) has no such appeal; but there is no cause for debate, since the status of Elga's argument turns on whether (C3) has such appeal, and we and Elga agree that there is no reason why intelligent folks should necessarily agree about the appeal of an undefended premise.  Nevertheless, from our point of view, Elga's original argument remains circular.

\vspace{-0.25cm}
\begin{acknowledgments} 
We thank Adam Elga for a very interesting and helpful email exchange and for his permission to include the summary of this exchange that appears in the final paragraph of Sec.~\ref{SEC:Conclusion}. We also thank Robert Audi, Jeremy Butterfield, Kris McDaniel, and Nicholas Teh for helpful advice. F.$\,$A.~acknowledges support from the Black Hole Initiative at Harvard University, which is funded through a grant from the John Templeton Foundation and the Gordon and Betty Moore Foundation.  A.$\,$H.$\,$G.'s~work was supported in part by the U.S.~Department of Energy under Contract No.~DE-SC0012567.  The views presented in this paper do not necessarily reflect those of any person or funding agency mentioned above.
\end{acknowledgments}

\appendix

\section{Generalizing Elga's Footnote 8}

In Section \ref{SEC:Bayes}, we defined $F$ to be the Bayesian update factor with which Al's credence in $H$ is multiplied when he acquires the new information ($HA \hbox{ or } TD$), that is, when he learns that either he is Al and the coin landed heads, or else he is the duplicate (Dup) and the coin landed tails.  In Eq.~\eqref{Eq:PAd}, we inverted the Bayesian update formula to determine $\Pinf(A)$, Al's credence in being Al, in terms of $F$ and $P_0(T)$.  [Recall that we are using the special notation $\Pinf(A)$ and $\Pinf(D)$ for Al's credence in being Al, or in being Dup, when expressed as a function of $F$ and $P_0(T) \equiv 1 - P_0(H)$.] This formula shows that if one assumes that $F = 1$, then one concludes that $\Pinf(A) = 1/2$.  Elga assumed that $F=1$, without justification in our opinion, and concluded that $\Pinf(A) = 1/2$.

Elga's demonstration that the claims (C1), (C2), and (C3) imply that $P(A) = P(D) = 1/2$ is given in his footnote 8, which does not use Bayes' theorem.  Since Elga's derivation is rather different from our derivation of Eq.~\eqref{Eq:PAd}, a reader could suspect that Elga's derivation might uncover more information than our Eq.~\eqref{Eq:PAd}.  Here we show that this is not the case: if Elga's derivation is generalized to allow $F$ to be arbitrary (rather than assuming that $F=1$), the final result is the same as Eq.~\eqref{Eq:PAd}.  To make this clear, we will go through the equations of Elga's footnote 8 step by step, but allowing for an arbitrary value of $F$.  At each step, Elga's equation can be obtained by setting $F=1$. We will indent the remainder of this paragraph to indicate that we are following Elga---for the most part using his language, to facilitate the comparison.
\begin{quote}
Elga begins by setting the left-hand sides of Eqs.~\eqref{EQN:C2} and \eqref{EQN:C3} equal to each other.  Using Eq.~\eqref{EQN:C3prime} instead of Eq.~\eqref{EQN:C3}, this gives
\ba \label{Eq:PHP}
P(H|HA \hbox{ or } TA)= \frac{1}{F}P(H|HA \hbox{ or } TD). 
\ea 
Rewriting Eq.~\eqref{Eq:PHP} using the definition of conditional probability, we obtain
\ba 
\dfrac{P(HA)}{P(HA\hbox{ or } TA)} = \dfrac{P(HA)}{F P(HA\hbox{ or }  TD)}.
\ea 
Some algebra then gets us that
\ba 
P(HA \hbox{ or } TA) =F P(HA \hbox{ or } TD).
\ea 
Since $HA$, $TA$, and $TD$ are all disjoint,
\ba 
\label{Eq:TA_T}
P(TA)=FP(TD)+(F-1)P(HA).
\ea
Since $P(TA)$ and $P(TD)$ add up to $P_0(T)$,
\begin{align}
\label{Eq:TA_TD}
P(TA)&=\dfrac{F}{1+F}P_0(T) -\dfrac{1-F}{1+F}\, P(HA),\\ P(TD)&=\dfrac{1}{1+F}P_0(T) +\dfrac{1-F}{1+F}\, P(HA).
\end{align}
Now set the left-hand sides of Eqs.~\eqref{EQN:C1} and \eqref{EQN:C2} (in the main text) equal to each other:
\ba
\label{EQN:PHA}
P_0(H) = P(HA|HA \hbox{ or } TA).
\ea
It follows that
\ba
\label{EQN:PTA}
P_0(T) = P(TA|HA \hbox{ or } TA).
\ea
Dividing the first equation [Eq.~(\ref{EQN:PHA})] by the second equation [Eq.~(\ref{EQN:PTA})], we obtain
\ba
\frac{P_0(H)}{P_0(T)} = \frac{P(HA| HA \hbox{ or } TA)}
{P(TA|HA \hbox{ or } TA)}. 
\ea 
Using the definition of conditional probability, we thus obtain
\ba 
\label{Eq:HA_TA_ratio}
\frac{P_0(H)}{P_0(T)}=\frac{P(HA)}{P(TA)}.
\ea 
Rearranging, we get that
\ba 
P(HA)= \frac{P_0(H)P(TA)}
     {P_0(T)},
\ea 
which in turn can be written as
\ba
P(HA)=\dfrac{F P_0(T)}{1-F+2F P_0(T)}P_0(H),
\ea 
where $P(TA)$ was replaced using Eq.~\eqref{Eq:TA_TD}. So, since $P(HD)$ and $P(HA)$ add up to $P_0(H)$ [and $P_0(H)+P_0(T)=1$],
\ba 
P(HA)=\dfrac{F P_0(T)}{1-F P_0(H)} P(HD).
\ea
Combining this with the fact that
\ba 
P(TA)=\dfrac{F P_0(T)}{1-F P_0(H)} P(TD),
\ea  
[as a result of Eqs.~\eqref{Eq:TA_T} and \eqref{Eq:HA_TA_ratio}], we have that
\ba\label{Eq:PAPD} P(HA \hbox{ or } TA)=\dfrac{F P_0(T)}{1-F P_0(H)} P(HD \hbox{ or } TD).
\ea 
\end{quote}

Elga's footnote 8 ends with the equation corresponding to Eq.~\eqref{Eq:PAPD}, but the argument can be spelled out by noting that since ($H \hbox{ or } T$) is true, Eq.~\eqref{Eq:PAPD} can be rewritten as
\ba\label{Eq:PAPD_E}
\Pinf(A)=\dfrac{F P_0(T)}{1-F P_0(H)} \Pinf(D).
\ea 
In this form we can see immediately that if we were to assume that $F=1$, we would recover Elga's result, that is, $\Pinf(A)=\Pinf(D)$.  More generally, however, since $\Pinf(D)+\Pinf(A)=1$, Eq.~\eqref{Eq:PAPD_E} implies that
\ba \label{Eq:PAd2}
\Pinf(A)=\dfrac{F P_0(T)}{1-F+2F P_0(T)},
\ea
in agreement with Eq.~\eqref{Eq:PAd}.



\begin{thebibliography}{99}

\bibitem[Birch(2013)]{birch_13}
Birch, J. (2013). 
On the `Simulation Argument' and Selective Scepticism.
\href{https://doi.org/10.1007/s10670-012-9400-9}
{{\it Erkenntnis} {\bf 78}, 95--107}.

\bibitem[Bostrom(2003)]{bostrom_03}
Bostrom, N. (2003).
Are We Living in a Computer Simulation?
\href{https://doi.org/10.1111/1467-9213.00309} 
{{\it Philosophical Quarterly} {\bf 53,} 243--255.}

\bibitem[Bradley(2007)]{bradley_07}
Bradley, D. (2007).
Bayesianism and Self-Locating Beliefs {\it or} Tom Bayes meets John Perry.
Doctoral Dissertation (Stanford University, CA, USA).

\bibitem[Bradley(2011)]{bradley_11}
Bradley, D. (2011). 
Confirmation in a Branching World: The Everett Interpretation and Sleeping Beauty.
\href{https://doi.org/10.1093/bjps/axq013}
{{\it British Journal for the Philosophy of Science} {\bf 62}, 323--342}.

\bibitem[Carroll and Sebens(2015)]{carroll+sebens_15}
Carroll, S. M. and Sebens, C. T.  (2015). 
Many Worlds, the Born Rule, and Self-Locating Uncertainty. 
\href{https://arxiv.org/abs/1405.7907}{arXiv e-prints:1405.7907 [gr-qc]}. Updated version of a paper in: \href{https://doi.org/10.1007/978-88-470-5217-8}{{\it Quantum Theory: A Two-Time Success Story. Yakir Aharonov Festschrift.}} D. C. Struppa and J. M. Tollaksen (eds.) Milan: Springer, (2014), pp.~157--169.


\bibitem[Dorr and Arntzenius(2017)]{dorr+arntzenius_17}
Dorr, C. and Arntzenius, F.  (2017). 
Self-Locating Priors and Cosmological Measures. In: \href{https://doi.org/10.1017/9781316535783}{{\it The Philosophy of Cosmology.}} K. Chamcham, J. Silk, J. D. Barrow, and S. Saunders (eds.) Cambridge: Cambridge University Press, pp.~396--428.


\bibitem[Elga(2004)]{elga_04}
Elga, A. (2004).
Defeating Dr. Evil with Self-Locating Belief.
\href{https://doi.org/10.1111/j.1933-1592.2004.tb00400.x}
{{\it Philosophy and Phenomenological Research} {\bf LXIX}(2), 383--396}.

\bibitem[Isaacs, Hawthorne, and Russell(2022)]{isaacs+al_21}
Isaacs, Y., Hawthorne, J., and Russell, J.~S. (2022). 
Multiple Universes and Self-Locating Evidence. \href{https://doi.org/10.1215/00318108-9743809}{{\it Philosophical Review}, {\bf 131}(3), 241--294}.

\bibitem[Meacham(2016)]{meacham_16}
Meacham, C.~J.~G.~(2016).
Ur-Priors, Conditionalization, and Ur-Prior Conditionalization.
\href{https://doi.org/10.3998/ergo.12405314.0003.017}
{{\it Ergo} {\bf 3}(17), 444--492}.


\bibitem[Sebens and Carroll(2018)]{sebens+carroll_18}
Sebens, C. T. and Carroll, S. M. (2018). 
Self-locating Uncertainty and the Origin of Probability in Everettian Quantum Mechanics.
\href{https://doi.org/10.1093/bjps/axw004}
{{\it British Journal for the Philosophy of Science} {\bf 69}, 25--74}.

\bibitem[Weatherson(2005)]{weatherson_05}
Weatherson, B. (2005).
Should we Respond to Evil with Indifference?
\href{https://doi.org/10.1111/j.1933-1592.2005.tb00417.x}
{{\it Philosophy and Phenomenological Research} {\bf LXX}(3), 613--635}.

\bibitem[Wilson(2017)]{wilson_17}
Wilson, A. (2017). 
The Quantum Doomsday
Argument.
\href{https://doi.org/10.1093/bjps/axv035}
{{\it British Journal for the Philosophy of Science} {\bf 68}, 597--615}.

\end{thebibliography}
\end{document}